# PROBING THE ERA OF GALAXY FORMATION VIA TEV GAMMA RAY ABSORPTION BY THE NEAR INFRARED EXTRAGALACTIC BACKGROUND

D. MACMINN* and J. R. PRIMACK

*Santa Cruz Institute for Particle Physics*
*University of California, Santa Cruz, CA 95064*

**Abstract.** We present models of the extragalactic background light (EBL) based on several scenarios of galaxy formation and evolution. We have treated galaxy formation with the Press-Schecter approximation for both cold dark matter (CDM) and cold+hot dark matter (CHDM) models, representing a moderate ($z_f \sim 3$) and a late ($z_f \sim 1$) era of galaxy formation respectively. Galaxy evolution has been treated by considering a variety of stellar types, different initial mass functions and star formation histories, and with an accounting of dust absorption and emission. We find that the dominant factor influencing the EBL is the epoch of galaxy formation. A recently proposed method for observing the EBL utilizing the absorption of $\sim 0.1$ to 10 TeV gamma-rays from active galactic nuclei (AGN) is shown to be capable of discriminating between different galaxy formation epochs. The one AGN viewed in TeV light, Mrk 421, does show some evidence for a cutoff above 3 TeV; based on the EBL models presented here, we suggest that this is due to extinction in the source. The large absorption predicted at energies $> 200$ GeV for sources at $z > 0.5$ indicates that observations of TeV gamma-ray bursts (GRB) would constrain or eliminate models in which the GRB sources lie at cosmological distances.

## 1. Introduction

We have developed an indirect test of the era of galaxy formation by modeling the extragalactic background light (EBL) in the near infrared portion of the spectrum—a domain difficult to observe directly due to galactic and zodiacal contamination (Matsuura et al. 1994, Mattila 1990). A new method for probing this spectrum has been suggested (Stecker et al. 1992) which involves the absorption of TeV $\gamma$-rays by the EBL. TeV $\gamma$-rays have recently been detected from Markarian 421 (Punch et al. 1992), an AGN at a redshift of $z = 0.031$. If several more distant TeV sources exist, it should be possible to observe for the first time the EBL, and thus constrain the timescale over which galaxies form. The potential of this technique is illustrated by the constraints it places upon an earlier claim for a detection of the EBL (Matsumoto et al. 1988).

It has been known for many years that high energy $\gamma$-rays from sources at cosmological distances will be absorbed by a diffuse background of long wavelength photons (Gould & Schréder 1967) through electron-positron pair

* Now at the University of Chicago, Dept. of Astronomy & Astrophysics.

production. This point had been of little interest, however, until the discovery of a class of high energy $\gamma$-ray emitters by the EGRET detector aboard the Compton Gamma Ray Observatory (Lin et al. 1992). These sources, all AGNs of various subclasses, were found to emit a substantial amount of energy in the GeV energy domain with a roughly $E^{-2}$ spectrum. While the physical description of the emission process is uncertain, it is possible that this spectrum could continue into the TeV range for some of these sources, a hypothesis given support by the detection of Mrk 421 at TeV energies. Mrk 421 is the nearest of the EGRET sources and should suffer from only limited EBL absorption of its $\gamma$-rays as we will show, but sources at greater distances should show evidence of absorption in the TeV domain which can serve as an indicator of the era of galaxy formtion.

## 2. Extragalactic Background Models

Models of the EBL require several inputs: a library of stellar spectra and a function describing the number distribution by mass of a stellar population (the initial mass function, IMF); a characterization of galactic classes (elliptical, spiral) and the evolution of their respective stellar populations (the star formation rate, SFR) and dust content; a function describing the formation of galaxies as a function of redshift; and assumptions about the geometry of the Universe—the density parameter ($\Omega_0$), the cosmological constant ($\Lambda$) and the Hubble constant ($H_0$). Such models have been created before (Partridge & Peebles 1967; Yoshii & Takahara 1988; Franceschini et al. 1994) with a variety of different assumptions, but each has a perhaps oversimplified model of galaxy formation. The conclusion reached by these earlier modelers is that the EBL proves to be a poor test of cosmology, meaning geometry, as the uncertainties in the IMF and SFR overshadow differences due to $\Omega_0$ and $H_0$. We concur that the EBL provides a poor test of geometry. But this by no means limits its usefulness as a probe of cosmology, for the dominant factor influencing the EBL is the era of galaxy formation, which in modern theories depends strongly on the nature of dark matter. For simplicity, we have chosen to consider here only those models with a flat ($k = 0$) geometry and $\Lambda = 0$, i.e. $\Omega = 1$; the arguments that the age of the Universe $t_0 \sim 13$ Gyr then require $H_0 \approx 50$ km s$^{-1}$ Mpc$^{-1}$. Considering also $\Omega_0 < 1$, especially with $\Lambda > 0$, would tend to increase differences due to the galaxy formation epoch as such models form galaxies earlier than $\Omega = 1$ CDM and much earlier than CHDM. A larger value of $H_0$, as has been suggested by Freedman et al. (1994), would not change the central conclusion of this work as the effects of the era of galaxy formation on the EBL are largely independent of $H_0$. The specific dark matter models used to characterize this era would likely be supplanted by a new generation of models; likely too, new models would differ in their predictions for the era of formation, and it is here that this proposed test acquires its value as a model independent probe.

Observational work has suggested several models for the IMF; we have considered the full range predicted by various authors (Scalo 1986; Basu & Rana 1992). We have assumed the SFR to have the functional form of a decaying exponential in time (Bruzual & Charlot 1993), with two classes of galaxies considered: ellipticals, composing 28% of the total and with a characteristic star formation time scale of 0.5 Gyr, and spirals, 72%, with a characteristic time scale of 6 Gyr. The instantaneous recycling approximation has been used to model the evolution of the gas content and metallicity within each galaxy (Tinsley 1980). Dust absorption has been treated in a fashion similar to Guiderdoni & Rocca-Volmerange (1987); the dust emission spectrum is taken from Désert et al. (1990), and is based on a three component model corresponding to the PAH molecules ($\sim$ 10 to 30$\mu$m), warm dust from active star forming regions (30 to 70$\mu$m), and cold "cirrus" dust (70 to 1000$\mu$m). Galaxy formation has been treated with the Press-Schecter approximation, which expresses the number density of galaxy halos of a given mass as a function of redshift based on the results of N-body simulations (Klypin et al. 1993). Two different cosmological models for structure formation have been considered: the cold dark matter (CDM) model ($\Omega_{CDM} = 0.9$, $\Omega_B = 0.1$, bias of 1.5), representing a moderately early era of galaxy formation ($1 < z_f < 3$), and cold+hot dark matter (CHDM) ($\Omega_{CDM} = 0.6$, $\Omega_\nu = 0.3$, $\Omega_B = 0.1$), representing a late era of galaxy formation ($0.2 < z_f < 1$). The latter is normalized to COBE and both are based on the simulations of Klypin et al. (1993). There is a difficulty in describing a galactic population in this manner in that the Press-Schecter function merely predicts the abundance of dark matter halos and makes no distinction between a cluster of galaxies and a single galaxy. One must therefore also assume a maximum galactic mass; we set the upper limit at $M_{\max} = 5 \times 10^{12}$ M$_\odot$ based on a comparison between the galactic luminosities obtained for a power law mass to light ratio and the normalization process described below, and we have checked that the results do not depend sensitively on $M_{\max}$.

The final parameter to be set is the normalization of the SFR, or the efficiency of star formation for a given galactic type. This can be directly tied to the local galaxy population through a comparison of model galactic luminosities with those of the nearby population. A number of studies have been made of the galaxy local luminosity function (LLF). Most recent are the work of Loveday et al. (1992) based on the Stromlo-APM southern sky survey (LEPM LLF), the LLF of Efstathiou et al. (1988) based on an average of several surveys (EEP LLF), and the LLF of DeLapparent et al. (1989) based on the CfA survey (LGH LLF). Each has characterized the luminosity distributions of the galaxies in the blue magnitude band as a Schecter function,

$$\phi(y)\,dy = \phi_* y^\alpha e^{-y}\,dy \quad \text{where } y = L/L_{*B}. \tag{1}$$

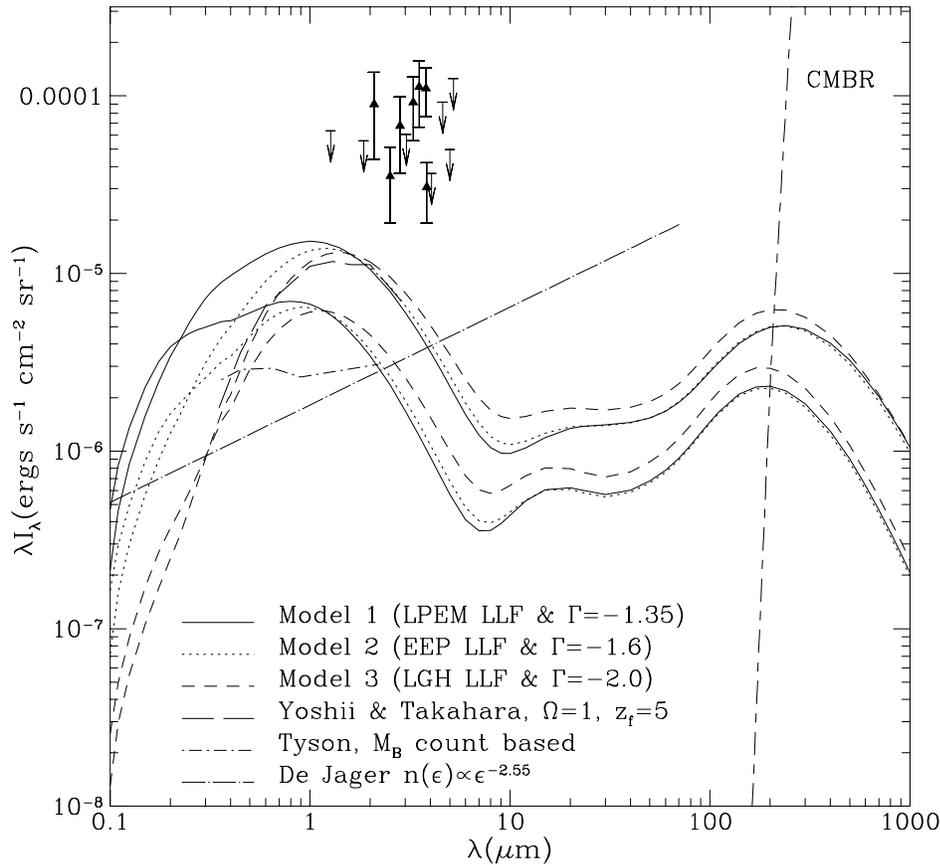

Fig. 1. The EBL at $z = 0$ for a variety of models. The upper set of curves is for the CDM model of galaxy formation, the lower set for CHDM. The number distribution of stars by mass was taken to be a power law of the form $N \propto m^{\Gamma-1}$; three values for $\Gamma$ were considered for each galaxy formation model. Two of those presented here represent the extreme values for both CDM and CHDM galaxy formation models, with Model 1 representing the bluest spectra and Model 3 the reddest; Model 2 is midway between these extremes. The near vertical line at the right represents the CMBR. Data of Yoshii & Takahara (1988) is taken from their model with cosmological parameters similar to those used herein; the Tyson (1990) data is a direct estimate of the EBL based solely on number counts of galaxies. The DeJager et al. (1994) model is based on a power law EBL fit to the possible absorption of Mrk 421 $\gamma$-rays shown in Fig. 4. Observations (triangles) and upper limits (arrows) are from Matsumoto et al. (1988); a similar experiment failed to confirm this measurement (Matsuura et al. 1994).

If one makes the assumption that the most massive galaxies are also the most luminous, one can compare the number densities of the LLF and the Press-Schecter function (at $z = 0$) and require that the average age model galaxies, at the present epoch, have the expected luminosities by deriving a function which expresses the SFR normalization as a function of galactic mass. The luminosity of the galaxies is thus normalized by forcing the model galaxies to duplicate the local population.

The model results are shown in Figure 1. Note the large separation between the galaxy formation model predictions, this will be seen for any combination of model parameters given that we include both early and late galaxy formation models and that we require that the present day model galaxies match the local population. The larger EBL flux predicted by the CDM model arises for several reasons: (1) stars have been contributing their light to the EBL for $\sim 2$ Gyr longer than in the CHDM case, (2) the initial burst of star formation in early-type galaxies has redshifted from the optical to the near IR, and (3) the galaxies are older at a given redshift and hence are composed of more evolved stars, producing a brighter flux in the red and near IR–see Figure 2. In addition to the parameter choices discussed above, we have also varied the other quantities not fixed by observational data to ensure that no large deviations were found.

There is some question as to the presence of a significant contribution from other sources than those considered here, particularly in the far IR ($\sim 100\mu$m) (although, as noted below, our primary interest is in the 1 to 10 $\mu$m region). The general nature of the galaxy models used herein is meant to represent both normal galaxies and the average properties of starburst and infrared luminous galaxies; the far IR flux predicted here is consistent with that of both Beichman & Helou (1991) and Hacking & Soifer (1991), each based on a luminosity function of extragalactic objects derived from the IRAS point source catalog and including quasar/AGN-like objects as well as infrared bright galaxies. Other sources not included in this catalog would have to be at high redshift or intrinsically dim; the former has no known population with which to be identified, and the latter would be of little effect. Given that we have not considered exotic IR producing mechanisms, however, the lowest curve in Figure 1 should be taken as a lower limit to the EBL.

## 3. TeV Gamma-Ray Absorption

The cross section for the absorption of $\gamma$-rays through $e^+e^-$ pair production is given by Gould & Schréder (1967); the total absorption is found by assum-

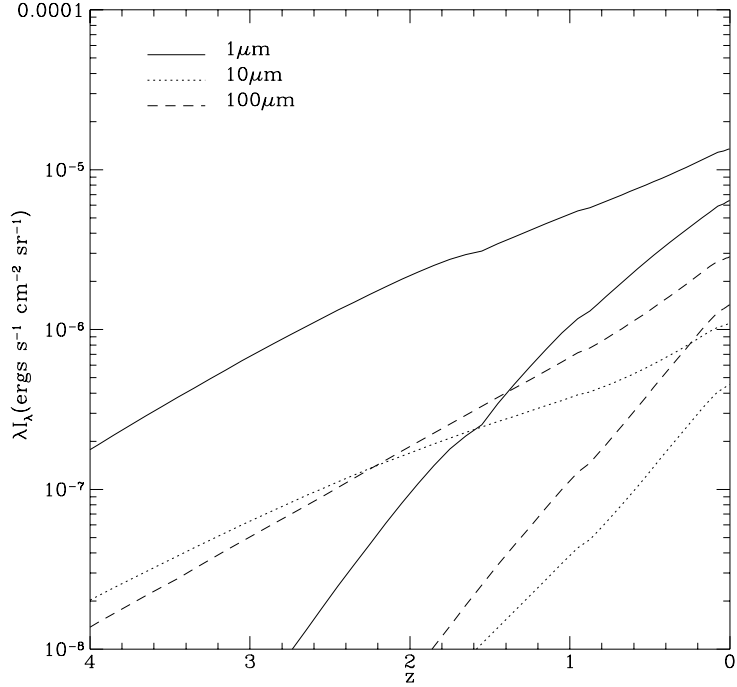

Fig. 2. The cumulative EBL flux at several different wavelengths plotted as a function of redshift of contributing sources. The upper three curves are for Model 2, CDM; the lower three for Model 2 CHDM. The flux at $1\mu$m is dominated by starlight, that at $10\mu$m by a mix of starlight and dust and at $100\mu$m by dust. Note that $\sim 30\%$ of the CDM model flux comes from sources at $z > 2$, and $\sim 50\%$ from $z > 1.0$. CHDM predicts only $\sim 1\%$ from $z > 2$, and $\sim 30\%$ from $z > 0.7$.

ing an exponential attenuation of a source's flux where the optical depth is given by

$$\tau(E) = \frac{c}{H_0} \int_0^{z_s} \int_0^2 \int_{2m_e^2c^4/Ex(1+z)^2}^{\infty} \frac{x}{2}\sqrt{1+z}\, n(\epsilon,z)\sigma(xE'\epsilon')\, dz\, dx\, d\epsilon \quad (2)$$

where $z_s$ is the redshift of the source; $E' = E(1+z)$ and $\epsilon' = \epsilon(1+z)$; $x = 1 - \cos\theta$, where $\theta$ is the encounter angle of the photons; $n(\epsilon, z)$ is given by the EBL models in Fig. 1; and $\sigma$ is the pair production cross section. This cross section is maximized when

$$\epsilon \approx \frac{1}{3}\left(\frac{1\text{ TeV}}{E}\right)\text{ eV} \quad (3)$$

where $\epsilon$ is the EBL photon energy and $E$ is the $\gamma$-ray energy. The photon number density for the EBL models drops off above 1 eV ($\sim 1\ \mu$m) and below 0.1 eV; the absorption effect due to the stellar component of the EBL will be most pronounced for γ-ray energies of 300 GeV to 3 TeV—see Figure 3. Above 10 TeV, it is likely that the dust-emitted component of the EBL makes the Universe mostly opaque, and above 100 TeV the cosmic microwave background completely absorbs any γ-rays from sources at cosmological distances. The predicted absorption at $E > 10$ TeV is similar to that suggested by Stecker et al. (1992) based on their EBL models, particularly for sources at $z > 0.1$. There is also an effect due to the redshift of the source. The absorption at some redshift along the path from the source will occur between a background photon of energy $\epsilon_0(1+z)$ and a $\gamma$-ray of energy $E_0(1+z)$, hence the lower cutoff in a source's spectrum will scale with $\sim (1+z)^{-2}$. For sources at $z \geq 0.5$ there is an additional effect from the galaxy formation models in that the magnitude of the EBL at large $z$ is quite small for cosmologies with late galaxy formation such as CHDM.

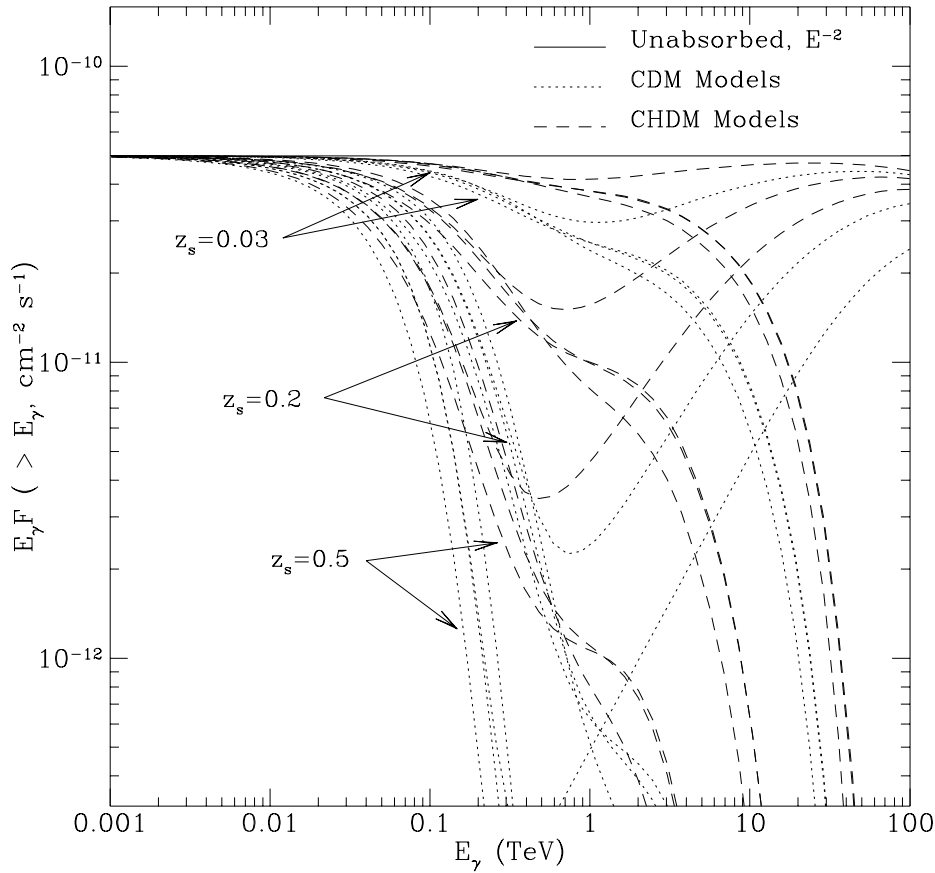

Fig. 3. The expected γ-ray flux from a source at three different redshifts ($z = 0.03, 0.2, 0.5$) with a spectrum given by $dN/dE = 5 \times 10^{-8} E_{GeV}^{-2}$ cm$^{-2}$ s$^{-1}$ GeV$^{-1}$ (typical of the EGRET detected AGNs). For each redshift, the three lower curves within each set represent Models 1, 2 and 3 from top to bottom. The upper curve in each set shows the absorption due to the stellar component of the EBL for the Model 2 parameters.

This last point, combined with the exponential dependence of $\gamma$-ray absorption upon the model predictions of the EBL, gives this test its potential power—provided that there are several sources which do possess a $E^{-2}$ $\gamma$-ray spectrum stretching into the TeV range, and that there are instruments with the sensitivity to detect fluxes on the order of a few events per cm$^2$ day. The one source known to emit TeV $\gamma$-rays, Mrk 421, may already constrain an earlier measurement of a diffuse infrared emission component of the sky which was attributed to the EBL—see Figure 4. The Mrk 421 spectrum also shows some evidence for a cutoff above 2 TeV; the size and sharpness of this cutoff is too large to explain by absorption with our physically motivated EBL models and may instead represent absorption intrinsic to the source. This explanation would suggest that the mechanism responsible for the production of the $\gamma$-rays was based on an electron Comptonization process (Dermer & Schlickeiser 1993; Sikora et al. 1994) rather than a proton based model (Mannheim 1993).

The EGRET team has observed at least 31 AGNs to date in the GeV energy range (Fitchel et al. 1994) of which 10 are at $z < 1$ and all with fluxes greater than that of Mrk 421. As of yet, only Mrk 421 has been seen in TeV light; the Whipple Observatory group has published upper limits on the TeV emission for two of the nearer GeV sources (3C 273 and 3C 279) (Fennell et al. 1993), but nothing conclusive can yet be said. EGRET has also observed several $\gamma$-ray bursts (GRB) in GeV light (Fitchel et al. 1994) with a comparable spectrum to the AGNs. It is possible that the process responsible for the GRBs may also produce a TeV flux, and this flux could be used to probe the extragalactic background as well. Given the strong absorption for $E_\gamma \geq 200$ GeV shown in Figure 3 for all of our models for sources at $z \geq 0.5$, the identification of any GRB at TeV energies would serve to constrain, or even eliminate, the hypothesis that the sources are at cosmological distances (see also the paper by Stecker and DeJager in these proceedings). Note also that the smallest absorption predicted here for any source should be taken as an upper limit to the TeV flux.

Observations at 100 GeV, below the cutoff expected for any model of the EBL, should help clarify whether the spectrum of a given source shows intrinsic absorption. The Whipple telescope is currently sensitive down to $\sim 500$ GeV. There are a number of instruments planned (and some funded) which can perhaps achieve lower thresholds as well as several other detectors to complement Whipple, chiefly the GLAST satellite, sensitive from 10 MeV up to $\sim 100$ GeV (GLAST workshop, 1994); the MILAGRO detector, operating above 300 GeV; and the Themis and Solar One arrays, operating at energies > 100 GeV (Rene Ong 1994, personal communication). If further sources are found to emit TeV $\gamma$-rays, the new detectors, combined with existing facilities, should be able to constrain many theories of galaxy formation. There are also a number of detectors operating at energies above 10 TeV—the CASA-MIA, CYGNUS, HEGRA, and TIBET arrays (Cronin et al. 1993); given the absorption of such $\gamma$-rays predicted in this work, it unlikely that such detectors will see *any* $\gamma$-rays from sources at cosmological distances.

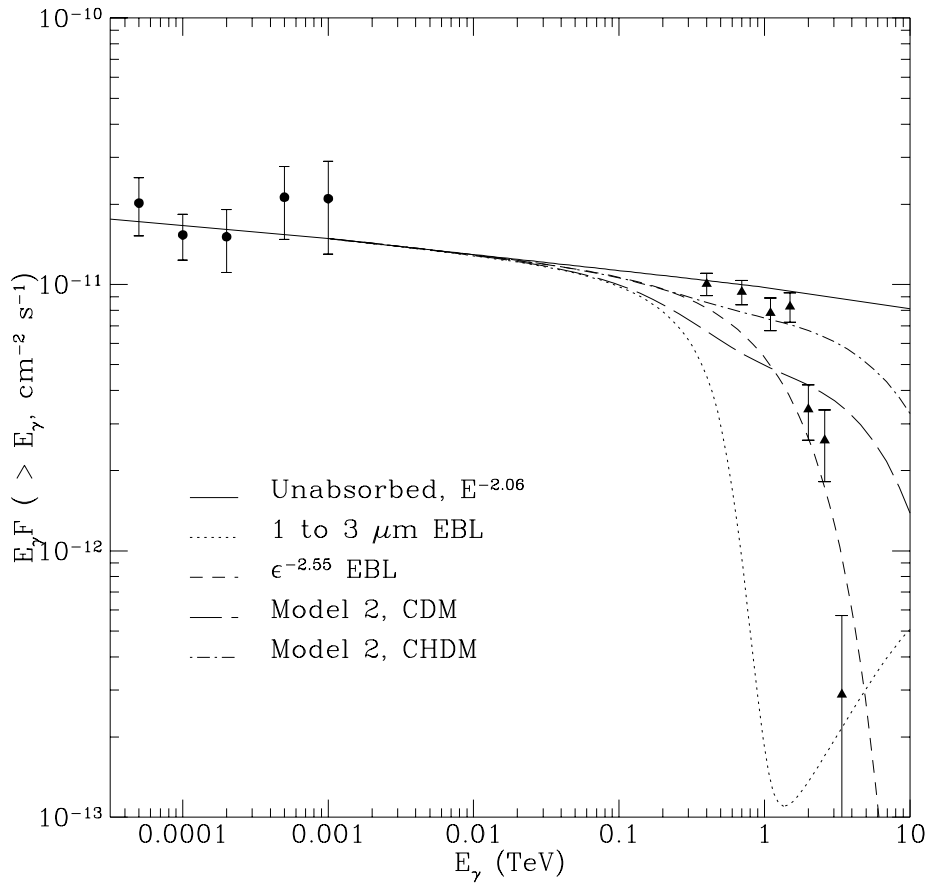

Fig. 4. The $\gamma$-ray flux from Mrk 421. Data is from the EGRET detector (Lin et al. 1992) (circles) and Whipple Observatory (Mohanty et al. 1993) (triangles). The Whipple data is somewhat uncertain due to difficulties with their energy calibrations (Richard Lamb and Trevor Weekes 1994, personal communications). The unabsorbed spectrum was determined by fitting a power law to the two data sets without assuming any absorption in the TeV domain; including absorption would lower the power somewhat (the EGRET data alone suggests an $E^{-1.96}$ spectrum) and bring both Model 2 curves into agreement with the first four Whipple points. The $\epsilon^{-2.55}$ is from the De Jager et al. (1994) model in Fig. 1. The 1 to 3 $\mu$m curve shows the absorption which would be expected if one interpets the measurement of Matsumoto et al. (1988) as a background of extragalactic origin; due to the sharp cutoff below 1 TeV, reasonable fits of the observed spectrum to this background are not possible.

## Acknowledgements

We have benefited from encouragement and helpful conversations with David Dorfan, Spencer Klein, Richard Lamb, Rene Ong, Floyd Stecker, Trevor Weekes and especially Anatoly Klypin. This research was supported by NSF REU funds and a grant from the DOE at UCSC.